\documentclass[aps,pra,twocolumn,10pt,showpacs,superscriptaddress]{revtex4-2}
\pdfoutput=1
\usepackage[utf8]{inputenc}
\usepackage[english]{babel}
\usepackage[T1]{fontenc}
\usepackage{amsmath}
\usepackage{hyperref}
\usepackage{natbib}
\setcitestyle{numbers}
\usepackage{bm}
\usepackage{amsfonts}
\usepackage{braket}
\usepackage{tabularx}
\usepackage[caption=false]{subfig}
\usepackage[ruled]{algorithm2e}

\usepackage{tikz}
\usepackage{lipsum}

\usepackage{xcolor}

\definecolor{dark-gray}{gray}{0.40}
\definecolor{quantumviolet}{HTML}{53257F} %%quantum
\definecolor{quantumlightviolet}{HTML}{A088B1}
\definecolor{quantumgreen}{HTML}{00826F}%{004E40}
\definecolor{quantumrose}{HTML}{EDB3FF} 
\definecolor{quantumdarkrose}{HTML}{F06292}
\definecolor{quantumturquoise}{HTML}{00C9AF}
\definecolor{quantumblue}{HTML}{85B1CC}
\definecolor{quantumdarkgray}{HTML}{4C4452}
\definecolor{quantumgray}{HTML}{555555}
\definecolor{black}{HTML}{000000}

\hypersetup{colorlinks=true,
            linkcolor=quantumgray,
            citecolor=quantumviolet,
            urlcolor=quantumdarkrose
}

\begin{document}

%\raggedbottom

\title{Photonic quantum generative adversarial networks for classical data}

\author{Tigran Sedrakyan} 
\affiliation{Quandela, 7 Rue Léonard de Vinci, 91300 Massy, France}
\affiliation{Sorbonne Université CNRS, LIP6, F-75005 Paris, France}
\author{Alexia Salavrakos}
\email[]{alexia.salavrakos@quandela.com}
\affiliation{Quandela, 7 Rue Léonard de Vinci, 91300 Massy, France}

\begin{abstract}
In generative learning, models are trained to produce new samples that follow the distribution of the target data. These models were historically difficult to train, until proposals such as Generative Adversarial Networks (GANs) emerged, where a generative and a discriminative model compete against each other in a minimax game. Quantum versions of the algorithm were since designed, both for the generation of classical and quantum data. While most work so far has focused on qubit-based architectures, in this article we present a quantum GAN based on linear optical circuits and Fock-space encoding, which makes it compatible with near-term photonic quantum computing. We demonstrate that the model can learn to generate images by training the model end-to-end experimentally on a single-photon quantum processor.
\end{abstract}

\maketitle

\section{Introduction}
Photonic quantum hardware represents one of the most promising paths for the realization of a quantum computer. There is a strong potential for scaling in the number of qubits \cite{scalablephotoniccomputer}, and various computation models and architectures have been designed for photonics \cite{raussendorf2001, Bartolucci2023}. Moreover, the possibility of demonstrating a near-term computational advantage in tasks such as boson sampling \cite{bosonsampling} makes it a noteworthy candidate for Noisy Intermediate-Scale Quantum (NISQ) technology \cite{Preskill_2018}. Even though building photonic circuits with large numbers of single photons and optical modes is challenging given current technology,  the rate of hardware advancement in the field is very encouraging \cite{pnr}.

Nevertheless, little effort has been dedicated so far to exploiting photonic-native architectures for machine learning tasks, i.e. architectures where the components are single photon sources, photon detectors, linear optical circuits with beam-splitters and phase-shifters, and where the problem is encoded in the Fock space. Some of the existing works concentrate on discriminative learning \cite{ganetal, Bartkiewicz_2020, maring2023}, but generative learning models remain mostly understudied \cite{generativereview, igans}, despite the great potential shown by classical generative models in recent years. While Fock-space based models do not consist of traditional qubits, they do exhibit quantum properties which could be harnessed in machine learning tasks.

In this work, we propose a quantum generative adversarial network (QGAN) where the generator network is a variational photonic quantum circuit \cite{Cerezo_2021}. We train our model on the MNIST \cite{deng2012mnist} dataset of handwritten digits in reduced dimension, using a patch-based image generation approach, in both ideal and noisy settings. We run the full training procedure as a physical experiment on Quandela's quantum processing unit \emph{Ascella} introduced in \cite{maring2023}, whose setup consists of a single-photon source connected to a integrated photonic chip and photon detectors. Our work is a proof-of-concept demonstration that photonic quantum adversarial models can be trained to generate classical data, and our results thus contribute to a growing body of literature on near-term quantum machine learning implementations.

\section{Background}

Generative models are designed to produce previously unseen data that follow certain patterns. Their training consists in feeding the model some target training samples that are representative of the desired outcome, and optimizing the model so that its outputs grow closer to these target samples. This corresponds to learning the underlying distribution from which the training samples are drawn, i.e. the data generating distribution. Generative models have a long-standing history, however, only recent advances in deep neural networks enabled the creation of deep generative models (DGMs), such as GANs, as well as variational autoencoders (VAEs), autoregressive and diffusion models \cite{gans, Kingma2014, Brown2020, pmlr-v37-sohl-dickstein15}. Among other factors, these advances were made possible due to training techniques and properties of deep neural networks \cite{deeplearningintro}.

\subsection{Classical GANs}
In this work, we focus on GANs as they are realizable on a relatively small scale and can be trained efficiently. First proposed by \cite{gans}, GANs marked an important milestone in the field of generative learning models. The primary concept of adversarial learning consists of two competing deep neural networks - the generator, often denoted as $G$, and the discriminator $D$. The generator accomplishes the task of data generation, by transforming the noise $\bm{z} \sim p_z(\bm{z})$ sampled from the latent space $\mathcal{Z}$ (also known as noise prior) into a fake data sample. Then, both generator and discriminator networks compete against each other in an adversarial zero-sum game, where the generator is trying to produce fake samples close to the real target samples drawn from the data generating distribution $\bm{x} \sim p_{data}(\bm{x})$, and the discriminator is trying to classify the real samples from the fake ones. The process is repeated iteratively until the generator starts producing realistic results, which may correspond to a Nash equilibrium being reached for the zero-sum game \cite{pmlr-v119-farnia20a}.

The learning process tries to maximize the loss value across the discriminator parameters, so that the discriminator is able to distinguish between the fake and real data. At the same time, it tries to minimize the loss over the generator parameters, so as to generate more realistic samples and confuse the discriminator. Mathematically, this is equivalent to a min-max optimization of a loss function $L(D, G)$ defined on the discriminator and generator models:
\begin{equation}
    \min_G \max_D L(D, G),
\end{equation}
where:
\begin{equation}
    \begin{aligned}
    L(D, G)  = \; & \mathbb{E}_{x \sim p_{\normalfont{data}}(\bm{x})}[\log(D(\bm{x}))] \\
     + \; & \mathbb{E}_{\bm{z}\sim p_z(\bm{z})}[\log(1 - D(G(\bm{z})))].
    \label{eq:gan-loss}
    \end{aligned}
\end{equation}
This can be expanded into the problem of maximizing two separate loss functions, the generator loss $L_G$ and the discriminator loss $L_D$:

\begin{equation}
\max_{\theta_G} L_G \; \text{  and  } \;  \max_{\theta_D} L_D,
\end{equation}
where:
\begin{equation}
    \begin{aligned}
        L_G & = \frac{1}{n} \sum_{i = 1}^{n}\log\left(D(G(\bm{z}_i))\right)
        \\  L_D & = \frac{1}{n} \sum_{i = 1}^{n}\left[ \log(D(\bm{x}_i))
        + \log(1 - D(G(\bm{z}_i)))\right],
    \label{eq:gan-final-loss}
    \end{aligned}
\end{equation}
with $n$ the number of training samples from the dataset, and $\bm{x}_i$ and $\bm{z}_i$ respectively the $i$-th real and noise samples. 

In practice, a batch of noise samples from the latent space is supplied to the generator. It produces a batch of results, which are then used for the discriminator. Simultaneously, the discriminator is supplied a batch of samples from the training dataset. The loss is therefore computed by averaging over the batch, which has the added advantage of stabilizing the learning process. For every learning iteration, optimization steps are performed first for the discriminator, and then for the generator, using gradient descent (or ascent), where the gradient is computed using backpropagation. The number of optimization steps $k$ for the discriminator depends on the specific use case. The full training algorithm is shown in the Supplemental Document in Section S1.

\subsection{Quantum GANs}

Quantum Generative Adversarial Networks (QGANs) were introduced in \cite{Lloyd_2018, Dallaire_Killoran_2018} as a quantum alternative to GANs. Rather than a single architecture, they consist of a set of concepts where at least one, if not both of the components - the generator and the discriminator - possess a certain degree of quantum capabilities. 

Moreover, not only the networks, but also the data or type of problem, can be quantum or classical, as is discussed in details in \cite{Lloyd_2018}. Additionally, the latent space can also be generated by a quantum source. References \cite{Rudolph_2022} and \cite{wallner2023towards} consider the alternative case where only the latent space of an otherwise fully classical GAN is generated quantumly, in qubit-based and photonic-native scenarios respectively. In \cite{Dallaire_Killoran_2018}, an early example for the generation of quantum data, i.e. the generation of a quantum state, is presented. The problem is defined so that in practice, the generator learns how to implement a CNOT quantum gate. A photonic example for the generation of quantum data was studied recently in \cite{Wang_2023}. For the generation of classical data with a QGAN, qubit-based models were developed in various works \cite{Zoufal_2019, Huang_2021, Situ_2020, romero2021}, and a model where the generator is based on the combination of a linear optical circuit and a neural network was very recently introduced in \cite{ma2024quantum}.

Here, we focus on the latter task of generating classical data, with photonic quantum circuits as a resource. In our scenario, the generator is a fully quantum variational circuit, while the discriminator is a fully classical discriminative neural network. The communication between both networks happens by obtaining classical samples from the output measurements of the generator, and feeding them to the discriminator, along with the target data. The rest of the training progresses as for classical GANs: the discriminator is optimized first, followed by the generator, iteratively.

\subsection{Image generation and patch-based approach}
The specific task we consider for our QGAN is to generate images. We use a patch-based approach in order to exploit only a small number of photons and modes, and thus make our scheme easier to execute experimentally.  The patch-based approach involves generating parts of larger images from separate quantum generators, which we call sub-generators, and combining or stacking the outputs together to obtain the full image. Such an approach has previously been shown to work on the $8\times8$ MNIST dataset in \cite{Huang_2021}.

\begin{figure}[t]
    \begin{center}
        \centering
        \includegraphics[width=0.35\textwidth]{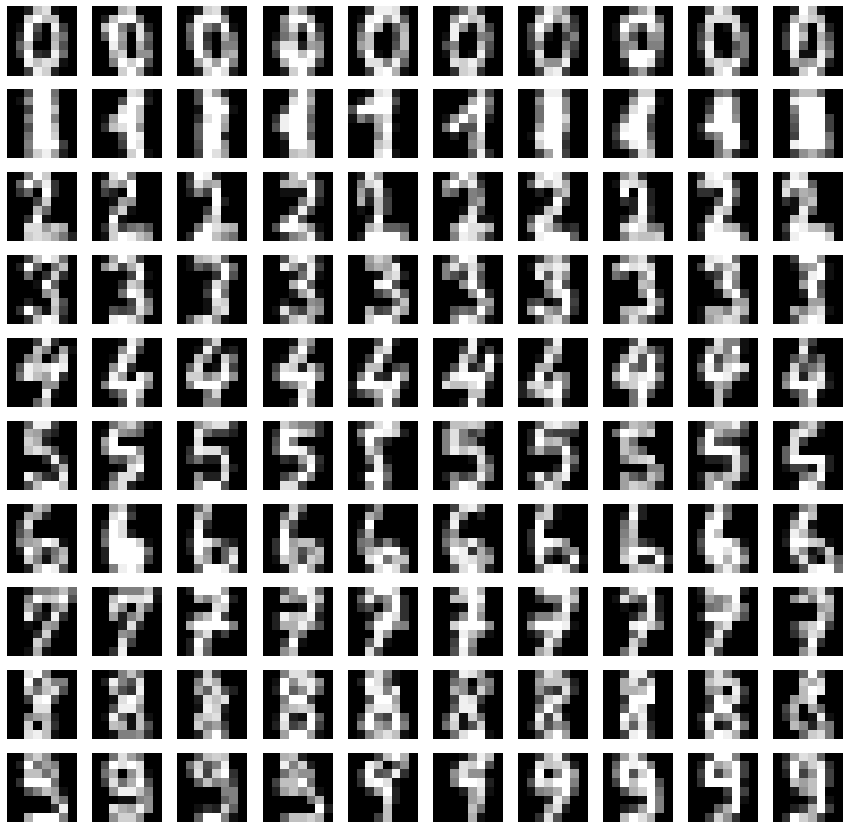}
        \qquad
    \end{center}
    \caption[Random entries from the $8\times8$ MNIST dataset]{Randomly sampled entries from the $8\times8$ MNIST dataset. Each row corresponds to a separate digit, sorted in increasing order 0-9.}
    \label{fig:mnist}
\end{figure}

We focus on this same dataset in our work. It is a downscaled version of the popular $28\times28$ MNIST dataset of handwritten digits, which was originally used for small-scale classical generative models. It consists of a collection of digit images ranging from 0 to 9, each image being of size $8\times8$. The dataset contains approximately $560$ entries for each digit, with $5621$ datapoints in total. Each datapoint consists of 64 pixels, and each pixel has continuous intensity value in the interval $[0, 1]$, with 0 being fully black pixels and 1 being fully white ones. Some examples, sampled randomly, are shown in Figure \ref{fig:mnist}. While the downscaling causes lower quality, the digits still closely resemble actual handwritten numbers and provide enough diversity for the tasks discussed in this work.

\begin{figure*}[t]
    \begin{center}
        \centering
        \includegraphics[width=0.9\textwidth]{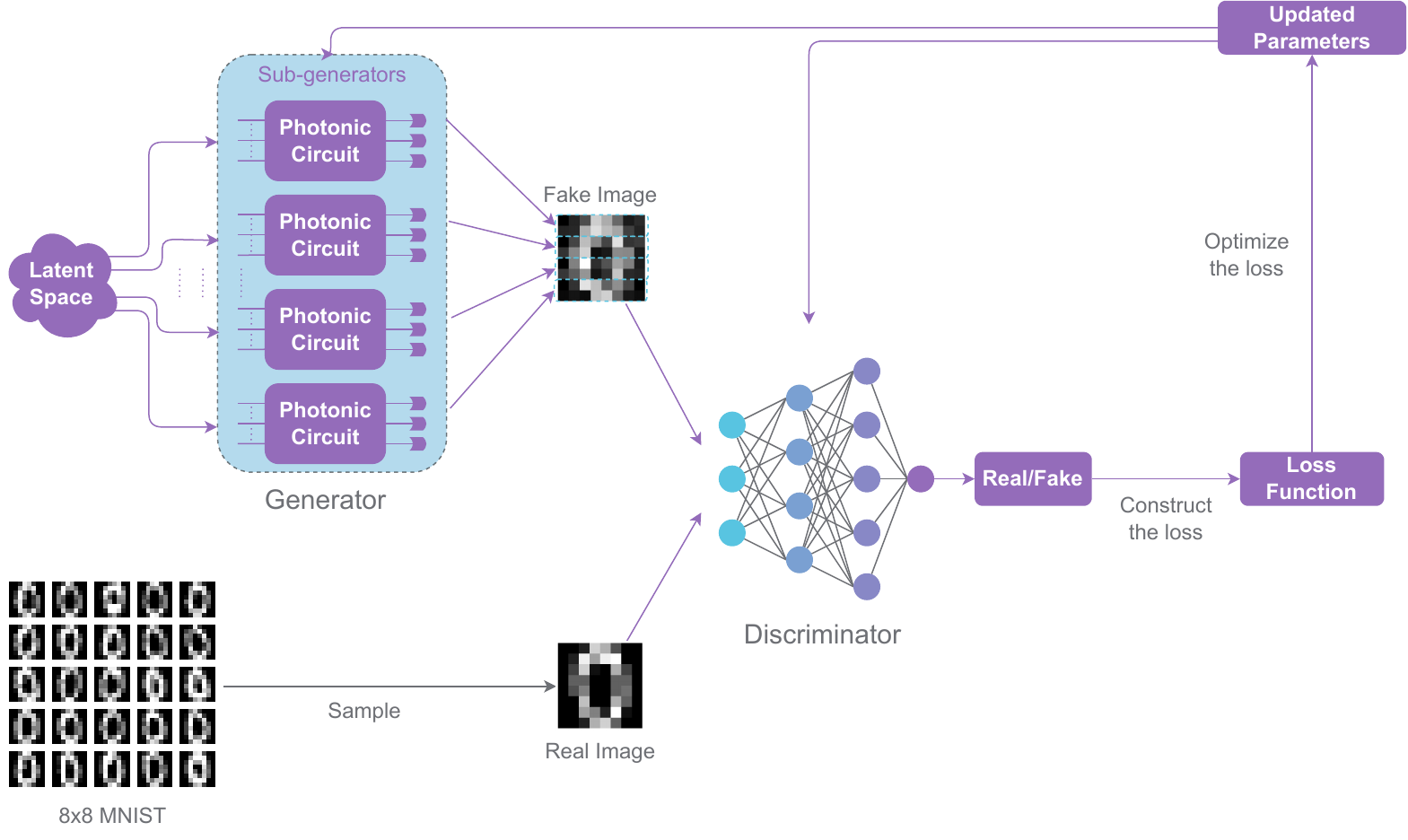}
        \qquad
    \end{center}
    \caption[Image generation with a photonic QGAN]{Proposal for image generation with a photonic QGAN. Noise from the latent space is fed into each sub-generator of the patch-based approach. These are variational photonic quantum circuits detailed  in Figure \ref{fig:qgan-circuits}. The output distribution of the sub-generators is mapped to image pixels, which are then recombined together to form a complete image, following the patch-based approach. The fake images are provided to the discriminator, along with the real images. The discriminator is a classical neural network, and classifies the image as real or fake. Based on these  results, the loss is constructed as per Equation (\ref{eq:gan-final-loss}). After optimizing the loss, the parameters of the generator and the discriminator are updated.}
    \label{fig:patchbased}
\end{figure*}

\section{A photonic QGAN}
\begin{figure*}[t]
    \begin{center} 
        \subfloat[Structure of a variational layer]{\centering
            \includegraphics[height=3.0cm]{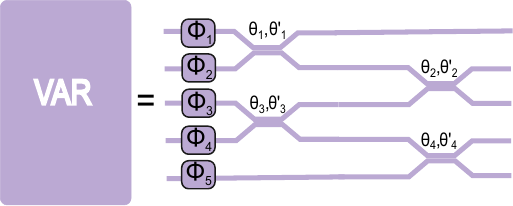}}
        \qquad
        \subfloat[Structure of an encoding layer]{\centering
            \includegraphics[height=3.0cm]{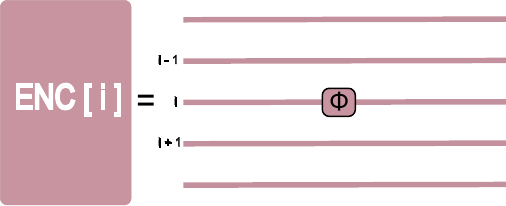}}
    \end{center}
    \caption{Structures of the variational layers and encoding or noise layers. Phase shifters are depicted by squares and beam-splitters by crossing between the modes. Parameters of the variational layers are trainable and parameters of the encoding layers are sampled from the latent space.}
    \label{fig:layer-structures}
\end{figure*}

We present our proposal for a photonic QGAN in Figure \ref{fig:patchbased}, where the quantum generator is implemented using linear optical variational quantum circuits. Having a photonic-native model means that the circuit ansatz consists of optical modes with parametrized phase shiters and beam-splitters, which is reflected in Figure \ref{fig:layer-structures}. In this framework, we consider as the input and output states of the circuit the Fock states of $n$ photons in $m$ modes, as in \cite{ganetal}. We denote an input Fock state as $\ket{\Vec{n}_{in}} = \ket{n^{in}_1, ..., n^{in}_m}$, where $n^{in}_i$ indicates the number of photons in mode $i$. Naturally, $\sum_i n^{in}_i = n$. Likewise, we can write an output Fock state as $\ket{\Vec{n}_{out}} = \ket{n^{out}_1, ..., n^{out}_m}$: they are detected as arrangements of photons in the output modes, which we denote $s 
 = (n^{out}_1, ..., n^{out}_m)$. If there is no photon loss, the $n^{out}_i$ sum to $n$ as well. 

An obvious way to design the quantum generator is to consider that one output state corresponds to one data sample, and to define a mapping between the Fock states and the space of the training data. As a simple example, let us imagine that we want to generate integers between $0$ and $100$, according to a certain target data distribution. We can then choose the number of modes $m$ and the number of photons $n$ such that there are at least $100$ possible output Fock states, and map each output state to an integer. In this scenario, one run of the quantum circuit produces one sample. This approach, which we could call sample-based, is used for instance in related work on photonic quantum circuit Born machines \cite{QCBMpaper}. 

However, while we are limited to small scale devices, such as the $12$ modes and $6$ photons of the \emph{Ascella} processor \cite{maring2023}, using this approach also means that we are restricted in the type of datasets that we can consider. Let us suppose that we aim to generate $8\times8$ MNIST digits, and that we use the patch-based approach as mentioned in the previous section with four patches, so that each circuit must generate images of 16 pixels at a time. If each pixel only had two intensity values (black or white), the dimension of the resulting space would be $2^{16}$.  This is already beyond what we could implement on \emph{Ascella}, and even more so if we considered the actual range of pixel intensities of the digit images. For this reason, and for the purpose of this work, we propose an alternative new mapping in the next section.

\subsection{A distribution-based mapping}\label{sec:mapping}
\begin{figure}[t]
    \begin{center}
        \centering
        \includegraphics[width=0.45\textwidth]{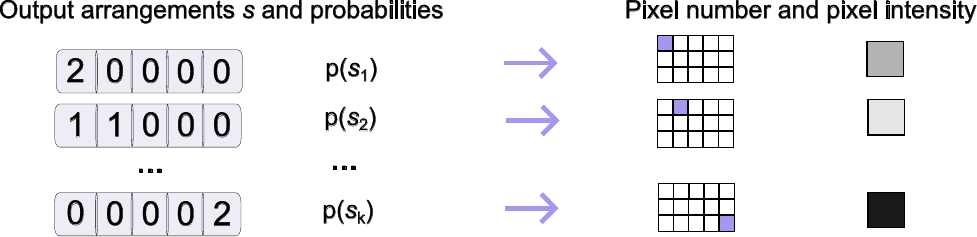}
        \qquad
    \end{center}
    \caption{Distribution-based mapping. Each output Fock state observed as arrangement $s = (n^{out}_1, ..., n^{out}_m)$ is mapped to a pixel number in the image, and the associated estimated probability is mapped to the intensity of the pixel.}
    \label{fig:mapping}
\end{figure}

In this approach, we compute the probability distribution on the output Fock states by performing several thousands of measurement shots at the end of the generator circuit. This discrete output distribution is then mapped to a discrete distribution on integers. If needed, binning may be performed so that several output Fock states correspond to the same integer. The index of a bin, i.e. the integer, corresponds to the location of a pixel on the image, while the probability of the bin corresponds to the pixel intensity, as shown in Figure \ref{fig:mapping}. This allows us to obtain continuous pixel intensity values. To cover their full range, the probability values are renormalized to the interval $[0, 1]$ using min-max normalization.

It is important to note that the number of possible output states of the generator does not always match the number of pixels necessary for the image or the patch, so the output distribution of each sub-generator may be trimmed equally on each tail, under the assumption that tails of distributions do not carry much information. 

When transforming Fock states to integers, while we can apply an arbitrary mapping scheme, it would intuitively make sense if photon arrangements physically close to each other in the device would correspond to integers that are close as well. We thus consider outputs with the most number of photons in the rightmost modes as closer to 0 in their integer mapping. Moreover, the larger the number of photons in the rightmost mode the smaller is the mapped integer. Correspondingly, states with a larger number of photons in the leftmost modes are considered further away from 0. Naturally, the number of available integers corresponds to the number of distinguishable states.

This approach is most efficient when photon number resolving (PNR) detectors  are available. One of the main advantages photon number resolution provides is that it allows us to observe a larger amount of output states for given values of $m$ and $n$. 
However, PNR detectors remain difficult to design with current technology. With threshold detectors, the only accessible values are binary - 0 (no photon) or 1 (click, i.e. presence of photons). Learning is of course still possible, but the mapping differs since several states with photon bunching are indistinguishable from each other. A sample mapping for 3 modes and 3 photons is shown in Table \ref{table:mapping}.%, both when PNR is available and when it is not.

\def\arraystretch{1.2}
\begin{table}[t]
    \centering
    \begin{tabular}{|c||c||c|c|}
        \hline
        \textbf{Integer} & \textbf{PNR} & \textbf{No PNR} (state)& \textbf{No PNR} (pattern) \\ 
        \hline
        0       & $\ket{0,0,3}$      & $\ket{0,0,3}$ & $\ket{0,0,\text{click}}$                 \\
        \hline
        1       & $\ket{0,1,2}$      & $\ket{0,3,0}$ & $\ket{0,\text{click},0}$                 \\
        \hline
        2       & $\ket{0,2,1}$      & $\ket{0,2,1}$,  $\ket{0,1,2}$ & $\ket{0,\text{click},\text{click}}$ \\
        \hline
        3       & $\ket{0,3,0}$      & $\ket{3,0,0}$  & $\ket{\text{click}, 0, 0}$                \\
        \hline
        4       & $\ket{1,0,2}$      & $\ket{2,0,1}$, $\ket{1,0,2}$ & $\ket{\text{click},0,\text{click}}$ \\
        \hline
        5       & $\ket{1,1,1}$      & $\ket{2,1,0}$, $\ket{1,2,0}$ & $\ket{\text{click},\text{click},0}$ \\
        \hline
        6       & $\ket{1,2,0}$      & $\ket{1,1,1}$    & $\ket{\text{click},\text{click},\text{click}}$              \\
        \hline
        7       & $\ket{2,0,1}$      & -  & -                             \\
        \hline
        8       & $\ket{2,1,0}$      & -       & -                        \\
        \hline
        9       & $\ket{3,0,0}$      & -      & -                         \\
        \hline
    \end{tabular}
    \caption[Fock state to integer mapping table]{Fock state to integer mapping table for a noiseless setup with 3 modes and 3 photons. The last column clarifies  the detection pattern without PNR.}
    \label{table:mapping}
\end{table}

This mapping assumes ideal conditions without photon loss. In noisy settings, the mapping does not need to be updated if PNR is available, since lossy states can be properly detected and filtered out of the final distribution with postselection. However, with threshold detectors, photon loss introduces an ambiguity in the output distribution and another mapping is necessary. In practice, lossless states with photon bunching cannot be distinguished from lossy states and they are both discarded in postselection. For the case of 3 photons with 3 modes, the number of output states reduces to one - $\ket{1,1,1}$. In such a situation, $m$ or $n$ must be increased to recover the necessary amount of integers for the image size.

\begin{figure*}[t]
    \begin{center} 
        \subfloat[Circuit setup A]{\centering
            \includegraphics[height=2.05cm]{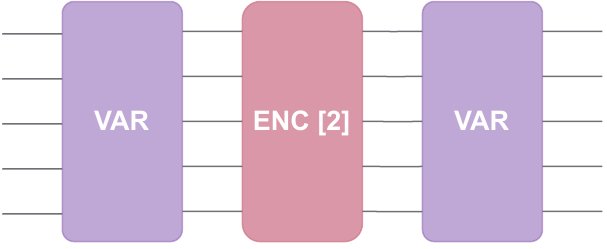}}
        \qquad
        \subfloat[Circuit setup B]{\centering
            \includegraphics[height=2.05cm]{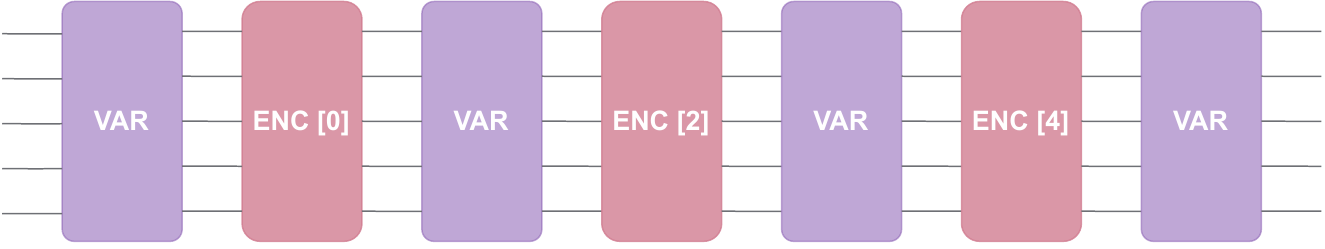}}
        \\
        \subfloat[Circuit setup C]{\centering
            \includegraphics[height=2cm]{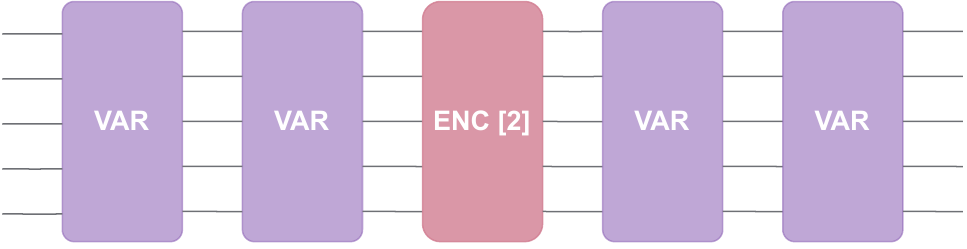}}
        \qquad
        \subfloat[Circuit setup D]{\centering
            \includegraphics[height=2cm]{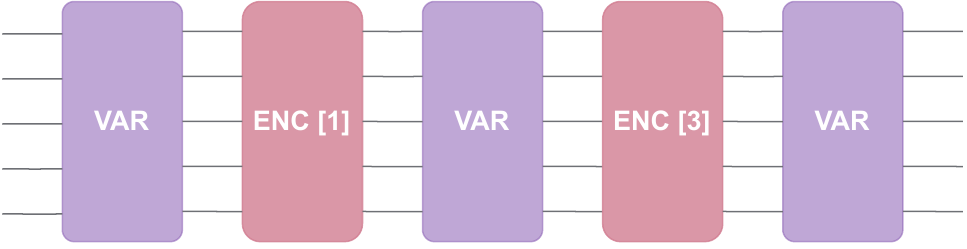}}
        \\
    \end{center}
    \caption[Sub-generator circuit structures used for photonic QGAN training]{Sub-generator circuit structures used for photonic QGAN training. Layer structures are shown in Figure \ref{fig:layer-structures}.}
    \label{fig:qgan-circuits}
\end{figure*}

\subsection{The ansatz}
In our patch-based approach, images are generated in horizontal patches by separate sub-generators and eventually stacked vertically to form the full image. Each sub-generator corresponds to a linear optical quantum circuit.
When designing the ansatz, we considered setups with  two quantum sub-generators, where each sub-generator generates patches of 32 pixels, as well as setups with four quantum sub-generators, where the patches contain 16 pixels.

In a given setup, all the sub-generators of the generator have the same structure. This structure consists of variational layers, and encoding or noise layers, as described in Figure \ref{fig:layer-structures}. The variational layers contain the parameters that are optimized during the training of the model and their structure is inspired from \cite{shankar2022variational, Shi2023}. The encoding or noise layers are used to introduce noise $z$ into the model (here sampled from a normal distribution). These encoding-reuploading layers consist of phase-shifters. They ensure that the resulting distribution over pixel intensities is different for each input noise sample, and that the model can thus generate a variety of data points, as well as generalize better. In general, noise-reuploading adds to the non-linearity of the input-output mapping, improving the diversity in the generated images and encouraging the model to learn patterns rather than memorizing them \cite{Chaudhary_2023, P_rez_Salinas_2020}.

We explored several structures for the sub-generator circuits. This structure can be adjusted by alternating the number and the arrangement of the variational and encoding layers. We display in Figure \ref{fig:qgan-circuits} the circuits that we found to be fairly efficient at solving our image generation task. %The primary difference in the structures is in the number, ordering and location of the layers. 
The smallest circuit only has one encoding layer, while the largest one has three. %The detailed structures of variational and encoding layers are shown in Figure \ref{fig:layer-structures}. 
The number of modes may vary compared to the circuits displayed in Figures \ref{fig:layer-structures} and \ref{fig:qgan-circuits}, but the general layer structure is preserved. It is important to note that considering the empiric nature of the findings, the circuit configurations are not guaranteed to be optimal, but are rather a heuristic combination of an educated guess and non-exhaustive search.

\subsection{Training and optimization}\label{trainingsec}
The training of photonic QGANs progresses similarly to the regular GANs, as in Algorithm S1. The classical discriminator is trained first for one step using backpropagation-enabled stochastic gradient ascent to maximize $L_D$, after which the quantum generator is trained for several steps of Simultaneous Perturbation Stochastic Approximation (SPSA)  \cite{705889} iterations, to optimize $L_G$. All the parameters are updated, and this process is repeated until the maximum number of training epochs is reached.

SPSA is an optimization technique based on a stochastic approximation of the gradient. Due to the fact that this approximation is an almost unbiased estimator of the gradient, convergence of the method is guaranteed under reasonably general conditions. The primary advantage of the SPSA algorithm lies in the amount of circuit evaluations necessary for approximating the gradient and its robustness to noise, including the noise induced by quantum sources \cite{wiedmann2023empirical,spsanoise}. SPSA requires only 2 evaluations, which allows cutting back on costly reconfiguration of linear optical gates and considerably reduces the duration of both simulations and experiments. Indeed, the number of training steps of the models scale only constantly (rather than linearly) with the number of parameters.

We initialize the parameters for SPSA in a way which allows the initial generator pseudo-gradients to be large enough for a successful kick-start to the optimization. In order to achieve this, parameters are initialized randomly, the initial gradients are computed and if the values are too small a reinitialization is performed. This parameter reinitialization is repeated until the starting pseudo-gradient values are in a desired range. Initialization performed in this way does not guarantee convergence, but in most cases allows the generator enough starting optimization momentum to be able to compete with the more exact gradient calculation of the classical discriminator, thus enabling a balanced training.

\subsection{Numerical experiments: ideal simulations and assessment}
\label{sec:idealsimulations}
We perform our numerical experiments using Quandela's software package \emph{Perceval} \cite{Heurtel2023percevalsoftware}, designed for linear optical quantum circuits. We include an abstract pseudocode in Supplemental Document in Section S1, and our Python code can be found in a companion Github repository \cite{photonicqgan}.

Ideal conditions for simulations assume a perfect single photon source, ideal components, no photon loss, perfect detectors and the absence of sampling errors.  First, we focus on the generation of digit "0" for the design of the model and the optimization of its hyperparameters. We then further study the best model for the generation of other digits, and for noisy simulations in the next sections. We include all details about the model optimization and hyperparameter search in the Supplemental Document in Section S2.

\begin{figure*}[t]
    \begin{center}
        \subfloat[Best training configurations for circuit setup A.]{\centering
            \includegraphics[width=0.45\textwidth]{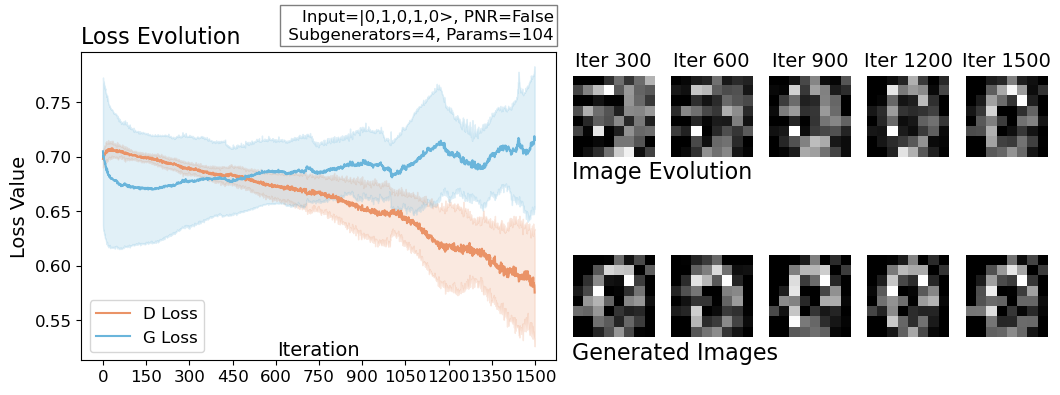}
            \hspace{1cm}
            \includegraphics[width=0.45\textwidth]{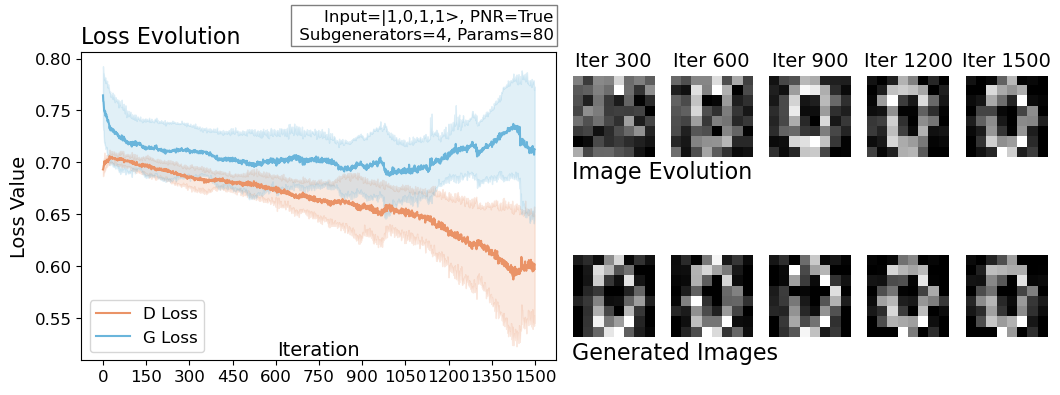}
            \label{fig:bestconf-A}
        }

        \subfloat[Best training configurations for circuit setup B.]{\centering
            \includegraphics[width=0.45\textwidth]{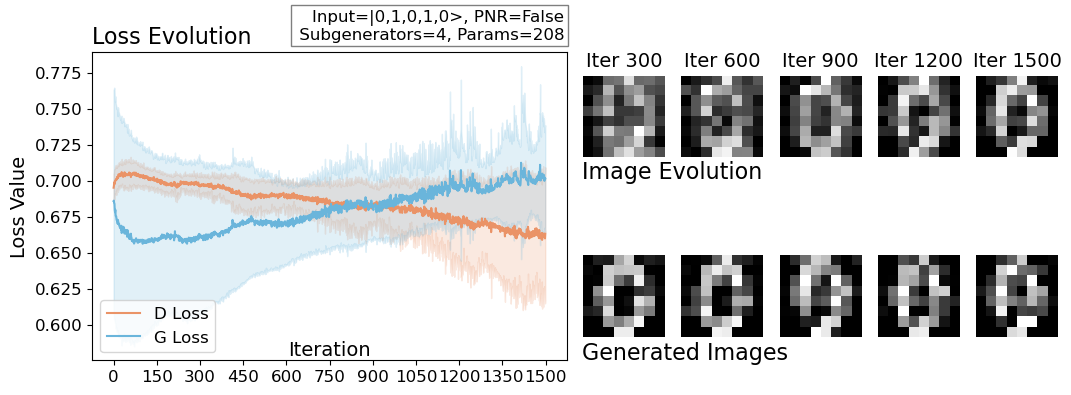}
            \hspace{1cm}
            \includegraphics[width=0.45\textwidth]{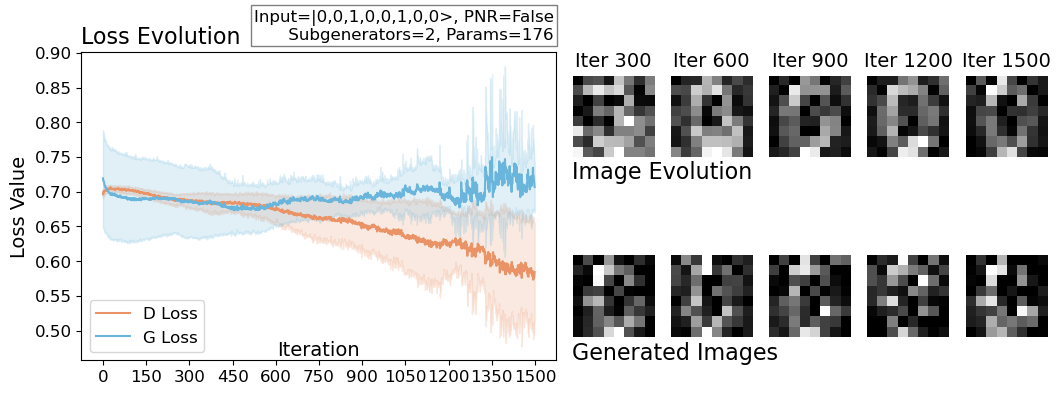}
            \label{fig:bestconf-B}
        }

        \subfloat[Best training configurations for circuit setup C.]{\centering
            \includegraphics[width=0.45\textwidth]{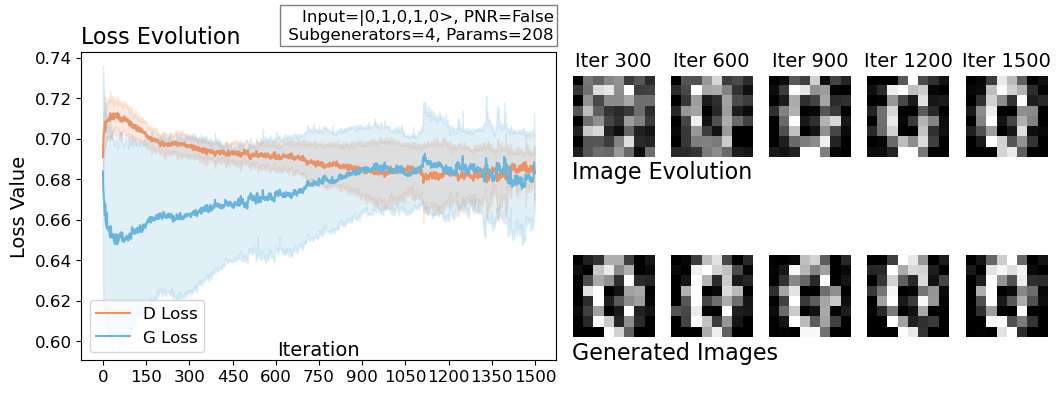}
            \hspace{1cm}
            \includegraphics[width=0.45\textwidth]{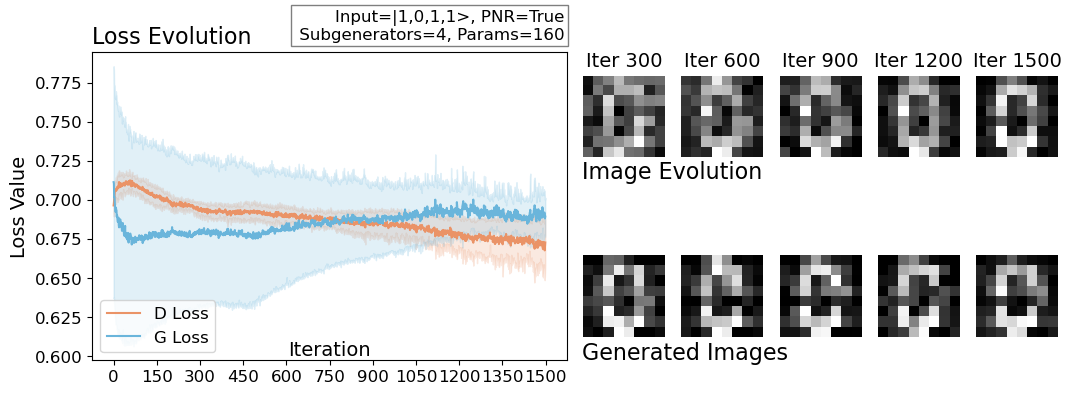}
            \label{fig:bestconf-C}
        }

        \subfloat[Best training configurations for circuit setup D.]{\centering
            \includegraphics[width=0.45\textwidth]{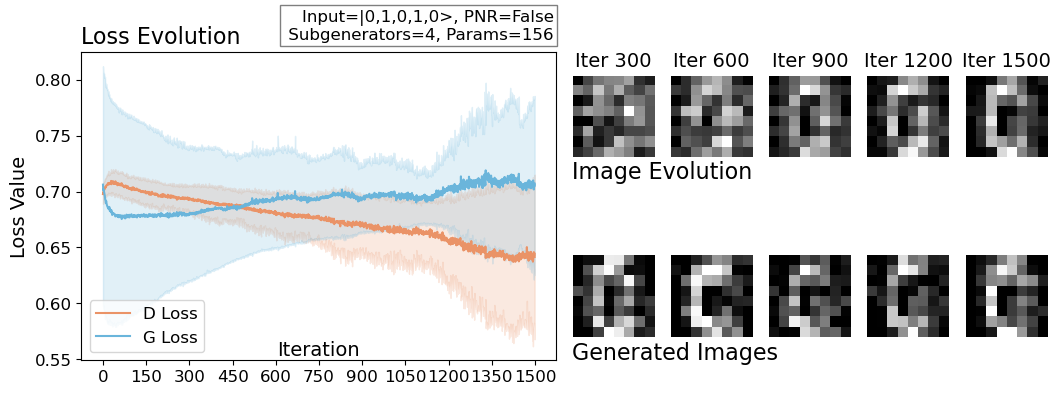}
            \hspace{1cm}
            \includegraphics[width=0.45\textwidth]{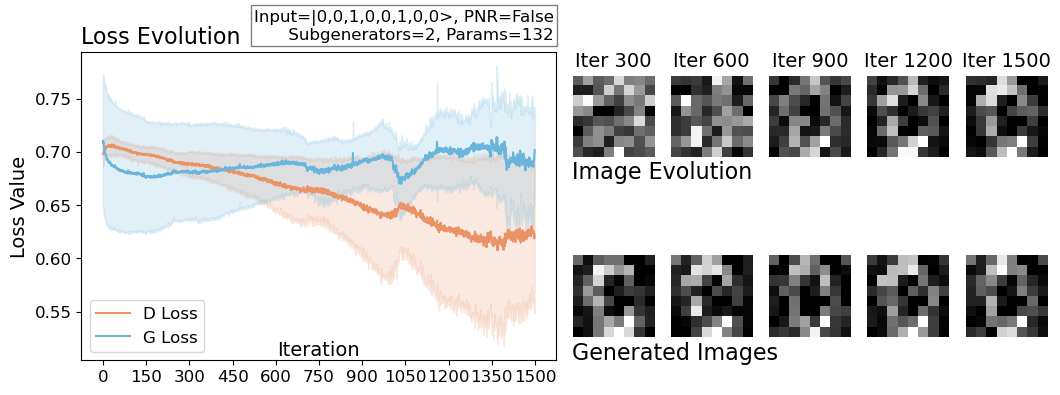}
            \label{fig:bestconf-D}
        }
    \end{center}
    \caption[Best training results]{Best training results for different \hyperref[fig:qgan-circuits]{circuit setups}. Each figure contains a plot with the evolution of the loss function, the evolution of the generated images as well as the final generated images of the trained model, and as a small infobox detailing the model hyperparameters. Figures on the left and right correspond to different sets of hyperparameters.}
    \label{fig:bestconf}
\end{figure*}

We present our results in Figure \ref{fig:bestconf}. The loss evolution plots describe how the values of the generator loss $L_G$ and the discriminator loss $L_D$ progress with the training epochs. For each configuration, we run 10 training instances, and we display the average over these runs as a bold line, and the standard deviation as a shaded area. We observe that even averaged over 10 runs, loss values have small constant fluctuations throughout the training, which is a strong characteristic of adversarial training. When the $L_D$ loss increases, the $L_G$ loss decreases, and vice versa. In the image evolution plots, we see that the models start with noisy outputs and that most of them produce quite realistic "0"s by the end of the training (iteration 1500). The generated image plot shows several samples from the trained model, after the last epoch has been completed.

It appears that there is no specific loss value where the training can be stopped. However, good models generally reach an equilibrium where $L_G$ and $L_D$ losses start to fluctuate around the same moving average, without increasing or decreasing further. 
This is clearly observed in Figure \ref{fig:bestconf-C} where both configurations converge to an equilibrium after around 1000 iterations. This equilibrium is centered around value close to $0.69$, which corresponds to $\log 2$. Taking a look at loss from Equation (\ref{eq:gan-final-loss}), it becomes clear that $L_G = -\log 2$ %(where the negative sign is due to the fact that a minimization of the inverse is performed instead of maximization)
, when $D(G(z)) = 1/2$. That is, when the discriminator prediction about whether the fake image is real is as good as a random guess. Despite the fact this equilibrium is reached around iteration 1000, we note that further training does improve the results. Therefore, cutting off the training once the losses reach the value of $\log 2$ is not necessarily a desirable strategy. % Rather, it makes sense to inspect the generated results using qualitative or quantitative metrics in order to be able to cutoff the training in due time. Quantitative assessment metrics are not discussed here, since the small scale of generated images allows for direct inspection of the results, as has been the case in many quantum generative learning works \cite{Huang_2021}.

% We refer the reader to Section S2 of the Supplemental Document for further analysis of the results and for hypotheses as to why some configurations perform better than others. Overall, the models from Figure \ref{fig:bestconf-C} behaved best and we select them for further simulations.

Loss does not constitute a good assessment metric, thus it makes sense to inspect the generated results using other qualitative or quantitative metrics. Important quantitative metrics in the assessment of generative models are the quality and the diversity of generated images. The quality measures the similarity to the original training data, while the diversity measures the variability between the generated images. Essentially, these metrics allow to evaluate a model's capability to learn well from data instead of memorizing it.

A common choice of measures of similarity and diversity are the Inception Score (IS) \cite{inceptionscore} and Fréchet Inception Distance (FID) \cite{fid}, which measure both the quality and the diversity of generated images. However, for a simple model working with lower dimensional images, such as ours, a simpler metric suffices. For this purpose, we chose the Structural Similarity Index Measure (SSIM) \cite{ssim}, which works by weighing in the comparative properties of luminance ($l$), contrast ($c$) and structure ($s$), defined as follows for patches $x$ and $y$ of images to be compared:
\begin{align*}
& l(x, y)=\frac{2 \mu_x \mu_y + c_1}{\mu_x^2 + \mu_y^2 + c_1}, \quad
c(x, y)=\frac{2 \sigma_x \sigma_y + c_2}{\sigma_x^2 + \sigma_y^2 + c_2}, \quad \\
& s(x, y)=\frac{\sigma_{xy} + c_3}{\sigma_x \sigma_y + c_3},   
\end{align*}
where $\mu$ and $\sigma$ are correspondingly the mean and the standard deviation of pixel intensity for the given patch, $\sigma_{xy}$ is the covariance between patches $x$ and $y$, and $c_1$, $c_2$ and $c_3$ are stabilizing constants, preventing a division by a small value. With these notions in mind SSIM is defined as:
\begin{align*}
\operatorname{SSIM}(x, y) & = l(x, y) \cdot c(x, y) \cdot s(x, y) \\
& =  \frac{\left(2 \mu_x \mu_y+c_1\right)\left(2 \sigma_{x y}+c_2\right)}{\left(\mu_x^2+\mu_y^2+c_1\right)\left(\sigma_x^2+\sigma_y^2+c_3\right)}.
\end{align*}
SSIM compares large images patch by patch and then averages the metrics across the patches, however for our case due to the small sizes of images, it compares them immediately. SSIM varies in the range $[-1, 1]$, with values close to $1$ indicating similarity, $0$ indicating absence of similarity and values close to $-1$ indicating anticorrelation between images.

In order to evaluate the models we compute the similarity score, which computes a pairwise SSIM of 500 random real images from the training dataset against 500 images generated by the quantum generator and then average across all the images. We also compute the the diversity score: first we compute the pairwise SSIM among the generated images, and again average across all images. This value is then subtracted from 1 to follow the logic of high diversity corresponding to more diverse (rather than more similar) images.

To have a baseline for comparison, we train a small GAN with a classical generator, which has approximately the same amount of parameters as the largest generator considered in this work (around 330). This model is one of the simplest possible generators with  only 1 small hidden layer, built in a way to allow for a fair comparison with quantum analog, with the training hyperparameters identical to those utilized for quantum generators. We note that a classical generator like this is heavily underparametrized and would not normally be used in a setting other than a comparative one, however it mimics well the behaviour of small quantum models discussed here.
\begin{figure}[h]
    \begin{center}
    \centering
    
            \includegraphics[width=0.4\textwidth]{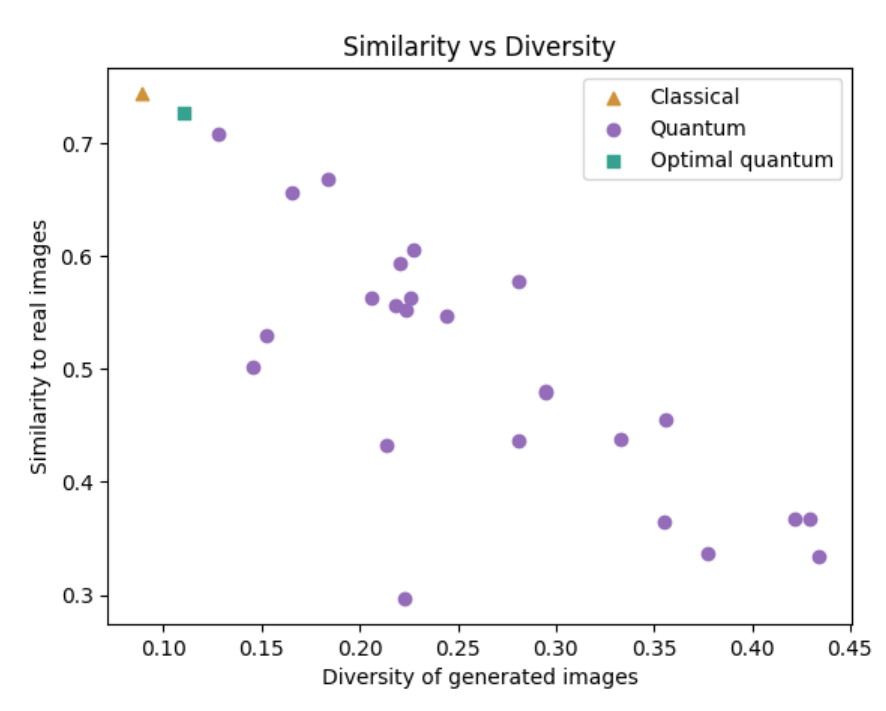}
        
    \end{center}
    \caption[Similarity vs Diversity]{Similarity plotted against diversity of generated images.}
    \label{fig:simdiv}
\end{figure}

We plot the similarity against the diversity for all the models tested, along with corresponding scores of the classical model on a scatter plot presented on Figure \ref{fig:simdiv}. Firstly, we note that diversity and similarity are anticorrelated for small models discussed here. In addition, the quantum model with the highest similarity (highlighted as a square, corresponding to Figure \ref{fig:bestconf-C}, left) has a performance comparable to that of the classical model (highlighted as a triangle), with slightly worse similarity and a slightly better diversity. Moreover, the quantum generator has around $2/3$ parameters of the classical one (208 vs 332).

Visual inspection confirms that indeed the results are of generally high quality for this model, albeit not very diverse. We thus choose this model for further testing, since it compares the best to the classical alternative. We refer the reader to Section S2 of the Supplemental Document for further analysis of the results and for hypotheses as to why some configurations perform better than others.

\begin{figure*}[t]
    \begin{center}
        \subfloat[Results for digit 1]{\centering
            \includegraphics[width=0.45\textwidth]{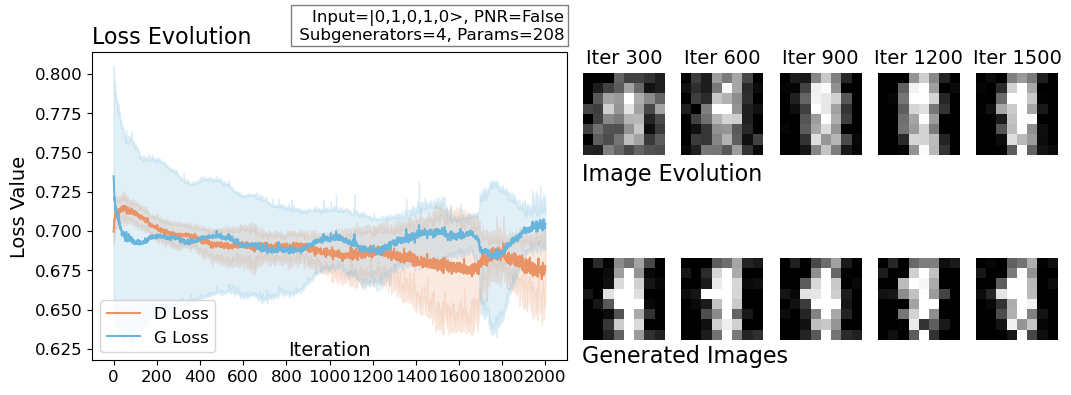}
        }
        % \qquad
        \hspace{1cm}
        \subfloat[Results for digit 3]{\centering
            \includegraphics[width=0.45\textwidth]{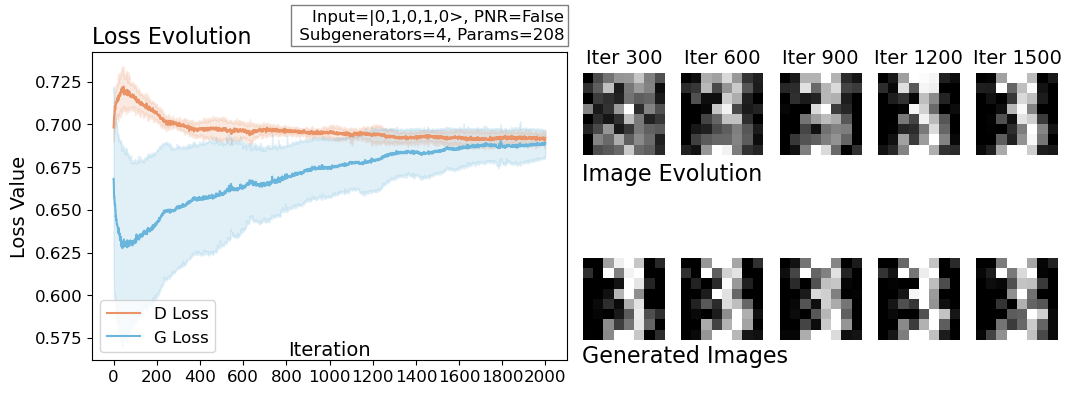}
        }
        \\
        \subfloat[Results for digit 5]{\centering
            \includegraphics[width=0.45\textwidth]{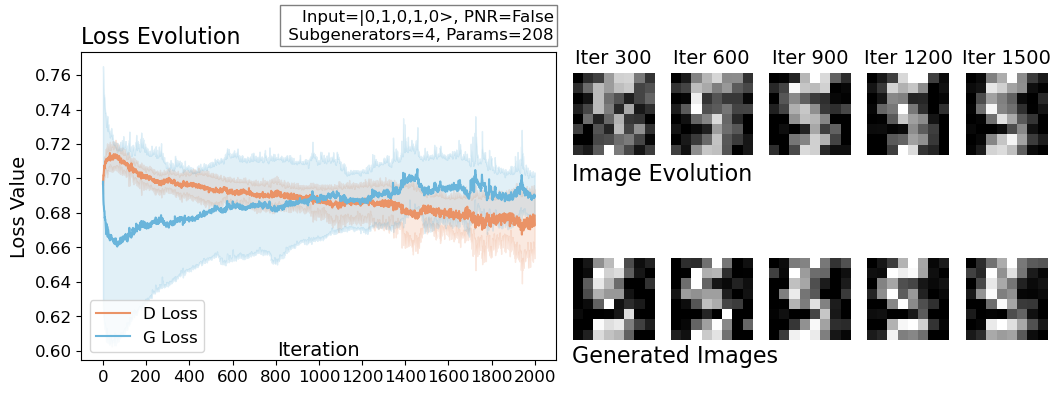}
        }
        % \qquad
        \hspace{1cm}
        \subfloat[Results for digit 9]{\centering
            \includegraphics[width=0.45\textwidth]{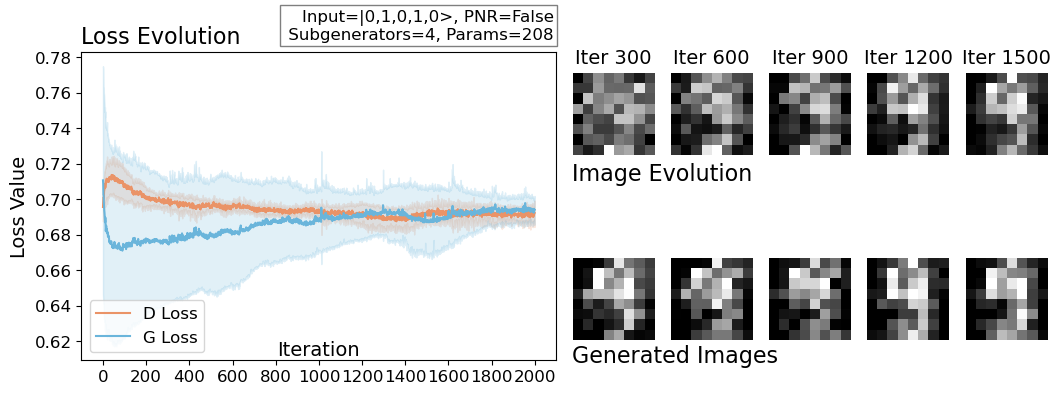}
        }
    \end{center}
    \caption[Training results for different digits]{Training results for different digits.}
    \label{fig:digits}
\end{figure*}

We apply the model from the left hand side of Figure \ref{fig:bestconf-C} to the generation of other digits. We present results for some digits in Figure \ref{fig:digits}, while the rest of the digits can be found in the Github respository \cite{photonicqgan}. An additional 500 iterations were employed here in order to properly assess the convergence for different digits.  
When comparing to target data, we see that the model performs fairly well for most of the digits, and each sampled digit has recognizable contours. Importantly, the model either converges or is bound to do so for all the digits, which is indicated by a narrowing of the standard deviation for the $L_G$ loss. While some digits may require more training iterations to reach an equilibrium state, we can assert that the model starts producing realistic results around iteration 1500. 

Note that, for simplicity, we restricted our model to generating one digit at a time (unlike classical GANs trained on MNIST). This is a similar approach to previous QGAN proposals such as \cite{Huang_2021}, but our model could be extended to a conditional QGAN in future work. It would then be able to generate all the digits depending on the input label, which could be supplied as a model input or encoded through an additional layer.

\subsection{Noisy simulations}

\begin{figure*}[t]
    \begin{center}
    \centering
    
            \includegraphics[width=0.45\textwidth]{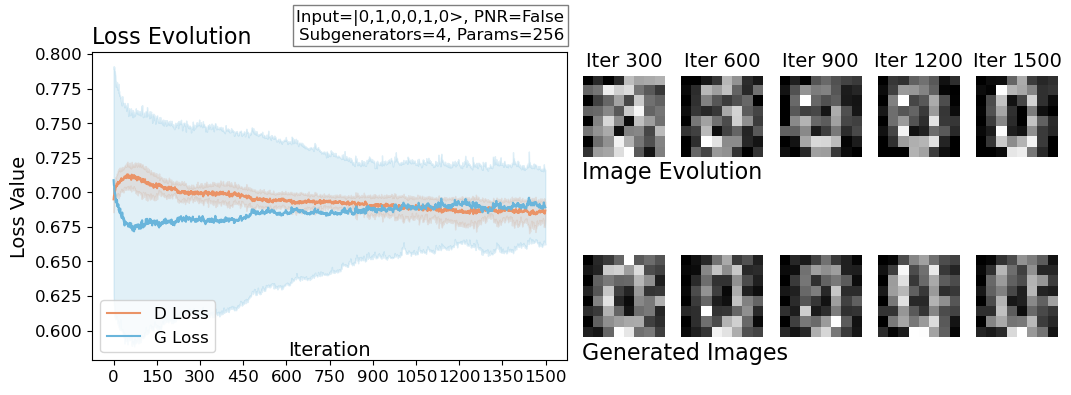}
            \hspace{1cm}
            \includegraphics[width=0.45\textwidth]{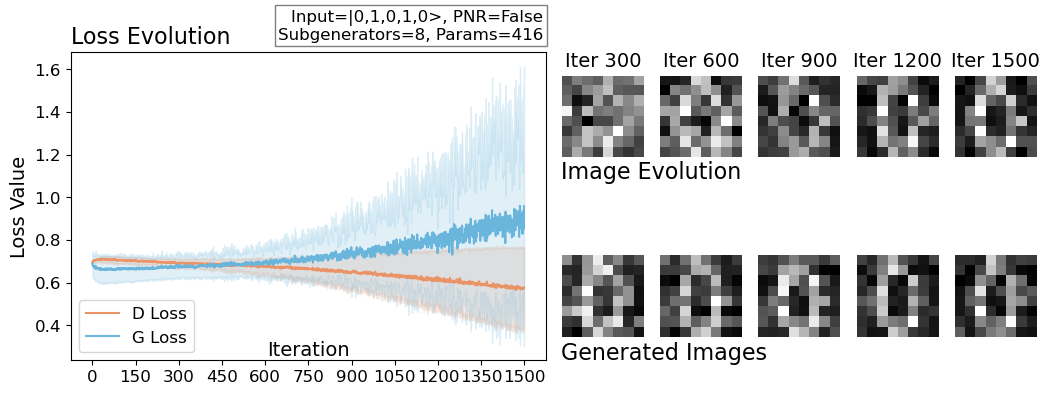}
        
    \end{center}
    \caption{Training results for noisy simulations. The model on the left has a higher number of modes, and the model on the right contains more sub-generators, which compensates for a smaller output space in the lossy case.}
    \label{fig:noisy}
\end{figure*}

We now use the same model for noisy simulations. The \emph{Perceval} package allows us to specify various parameters such as the emission probability of the source, the photon loss regime, and photon distinguishability. 
We set indistinguishability and photon loss to 0.92 each, so as to closely mimic the actual conditions on the \emph{Ascella} processor as it was presented in \cite{maring2023}.

In addition to the imperfect source and losses throughout the circuit, noisy simulations also introduce sampling error. For ideal simulations the distribution of the output with exact probabilities is directly available in \emph{Perceval}. However, in this case the distribution is obtained by  collecting $10^5$ measurement shots and postselecting lossy outputs. A discrete distribution obtained in such a way gets closer to the exact distribution with an increasing number of measurement shots. However, in an experimental setting, collecting a high number of shots requires precious time, and $10^5$ shots were found to be an optimal compromise for the accuracy/training time trade-off.

As discussed in Section \ref{sec:mapping}, the lack of PNR detectors combined with photon loss shrink the size of the output space. This requires us to change some hyperparameters and we display two strategies in Figure \ref{fig:noisy}.

Clearly, the added noise considerably slows down the convergence. While the noiseless version reaches an equilibrium around iteration 900 on average, for the noisy version on the right of Figure \ref{fig:noisy}, even 1500 iterations are not enough. This version, which has more sub-generators, shows a diverging learning trend. This might be caused by the increased number of parameters, making it harder to train this model under the predefined learning rate restrictions (see Supplemental Document, Section S2). Nevertheless, for the model on the left which has less sub-generators but a higher number of modes, one can see how the standard deviation of the generator decreases throughout the training, indicating that it is likely to eventually converge. 
Moreover, the generated results are satisfactory when applying a manual check. 
Results comprehensible to the human eye are available by iteration 600 for the ideal version, while the noisy version requires more than 1000.

Overall, training in the presence of noise and sampling errors comparable to those of a real quantum processor is still viable, albeit slower. Techniques for quantum error mitigation \cite{QEMreview} might improve the results of noisy simulations, and could be explored in future work.

\subsection{Physical experiment}

Based on the insights gathered from our simulations, we run an experiment on Quandela's processor \emph{Ascella}, with our best performing model. The experimental setup is the following: a gated InGaAs quantum dot placed inside a micropillar cavity \cite{Somaschi_2016} is excited by a laser as in the scheme of \cite{Thomas2021} to produce single photons on demand. A demultiplexer converts the train of produced photons into single photons arriving simultaneously on the chip, which is a 12-mode interferometer designed according to the Clements scheme \cite{Clements_2016}. The phase shifters on the chip are tuned thermo-optically in order to apply the intended operations, and this process is optimised following the characterization procedure of \cite{Fyrillas2024}. The photons then enter superconducting nanowire single-photon detectors (SNSPD) detectors, which are threshold detectors. As mentioned in the previous section, the setup is affected by photon loss and distinguishability.

\begin{figure}[h]
    \begin{center}
    \centering
    
            \includegraphics[width=0.45\textwidth]{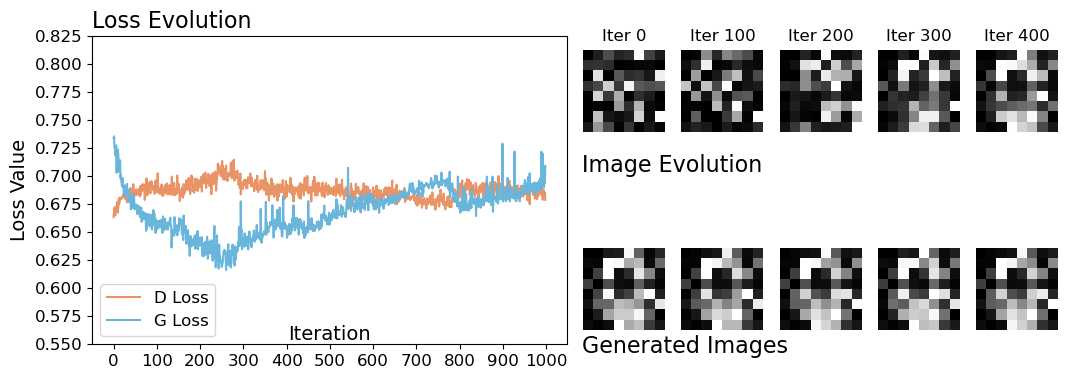}
        
    \end{center}
    \caption[Training results for the QPU experiment]{Training results for the experiment on the \emph{Ascella} quantum processor.}
    \label{fig:qpu}
\end{figure}

The circuit that corresponds to our best performing model is of reasonable depth, making it compatible with the \emph{Ascella} chip, which is naturally reused for each sub-generator. In order to make the duration of the experiment tractable, we decrease the total number of training iterations to 1000, with 3 SPSA steps for the generator at each iteration, which results in a total of 3000 SPSA steps. The results for one training instance are displayed in Figure \ref{fig:qpu}.

While the training is slow, following the trend of noisy simulations, the results are promising. Importantly, we observe that the model is learning, with the loss functions behaving as they should, tending towards the equilibrium value, and results improving throughout the training. A point that would require improvements in future experiments is that the generated images do not present a lot of diversity. There are only minor differences between the "0"s which might be improved by adding more encoding phase-shifters into the ansatz. Nevertheless, the generated "0"s are of fairly good quality, with characteristic contours of the digit, demonstrating the experimental feasibility of photonic QGANs.

\section{Discussion and future work}
In this work, we showed that photonic quantum circuits can work as a key component in a generative learning pipeline to produce images. This is in contrast to previous work where linear optical circuits were used for smaller-scale tasks such as latent space generation, without going as far as using a photonic generator. To the best of our knowledge, our experiment is the first demonstration of a photonic GAN with a fully quantum generator for classical data. Additionally, with the aim of transparency and collaboration within the quantum machine learning community, we make our code available online at \cite{photonicqgan}.

Most of the QGAN literature concentrates on smaller scale models for generation of lower resolution images, such as $3\times3$ bars-and-stripes. However, the task of generating larger images is tractable with currently available hardware if the right approach is used -- such as our distribution-based mapping. This highlights the importance of looking for alternative strategies when working with near-term quantum hardware. Using our mapping along with noise reuploading gave us greater flexibility, and in combination with patch-based learning the model could generate higher resolutions images. 

In terms of scaling, we point out that our approach requires the approximation of the output probability distribution and may thus have a high sampling cost, like many quantum machine learning models. Photonics at least have the benefit of providing fast sampling which improves runtime, a strength which will become more apparent as levels of photon loss decrease with the advancement of the technology. However, beyond a certain dimension of the problem, it might be necessary to modify the approach, for instance using our model together with a VAE that transforms the data.

We note that, while the size of the Fock space grows exponentially with the number of modes and photons, it is also directly linked to the dimension of the target data, given our mapping. Many problems of interest might thus be reachable with a reasonably sized device. When comparing the native photonic approach to the qubit-based approach of \cite{Huang_2021}, we see how in terms of resources utilised, on photonic platforms it may be advantageous to use the Fock space over the Hilbert space. As we saw above, a model can be built which uses 5 modes and 2 photons for the generation of 8x8 images, with the input state $\ket{0, 1, 0, 1, 0}$, as compared to 8 modes and 4 photons, which would be needed in dual-rail encoding to have qubits as in \cite{Huang_2021}.

Concerning the optimization, we note that SPSA seems to work surprisingly well on variational linear optical circuits. Especially in the noisy experimental setup, SPSA allowed  for stable learning at a constant cost, albeit requiring more iterations to converge. Fine-tuning the optimization and the model in general was an important aspect of the work. Future work may concentrate on improving the convergence rates of optimization techniques for photonic platforms, which can be used directly to improve the performance of our model. Indeed, only few works so far have focused on exploring methods like parameter-shift rules \cite{Schuld_2019} in the context of quantum photonics \cite{defelice2024, pappalardo2024, facelli2024}. As mentioned in the text, other improvements to this work include the extension of the model to a conditional QGAN, as well as the integration of error mitigation techniques for sources of noise such as photon loss and distinguishability \cite{Endo_2018}.

Overall, our QGAN implementation is quite flexible and modular, so the model can be further explored for generation of other types classical data, by fine-tuning parameters and introducing some changes to the circuit structure. For instance, a generator may consist of sub-generators with different structures, depending on the task at hand. We could also consider a quantum discriminator and training data based on a quantum source, for the generation of quantum states, for instance. Additionally, the source of the latent space could also be changed to be quantum, by using another boson-sampling-based circuit as a source of noise, as in \cite{wallner2023towards}.

In terms of applications, it is not always clear for which problems a quantum model will perform best -- a recent study showed that several popular quantum classifiers do not outperform their classical counterparts on standard datasets \cite{bowles2024better}. A solution that is often proposed is to focus on quantum data instead, e.g., to generate quantum states. However, problems containing classical data are overall much more common. There, the concept of inductive bias proves very helpful to follow, requiring that the model contain some information about the structure of the problem at hand. For this reason, high-energy physics and quantum chemistry are often cited as areas where quantum models might be most useful on classical data. For example, in \cite{Bermot_2023}, a QGAN is trained to detect anomalous events that cannot be described by the standard model; in \cite{rehm2024quantumgenerativeadversarialnetwork} a QGAN was used to generate calorimeter shower images; and in \cite{vakili2024quantumcomputingenhancedalgorithmunveils, Kao_2023}, QGANs and QCBMs were used for the generation of new molecules.

We saw that our best performing generator achieves similar performance to its classical counterpart for a lower number of trainable parameters. Such benchmarks can be interesting to focus on, as an alternative to speedups. In the same spirit, the energy consumption of quantum versus classical computers are currently being explored \cite{quantumenergy}.

%As is often the case in quantum machine learning, it is not always clear if a quantum model will perform better than a classical one and on which datasets that would be the case \cite{bowles2024better}, although interesting alternative benchmarks such as the energy consumption of quantum versus classical computers are being explored \cite{quantumenergy}. Tasks based on quantum data clearly require a quantum model -- for example if the generator learns how to implement a quantum gate or generate specific quantum  states. For classical data on the other hand, a deeper investigation is required, likely related to the identification of inductive biases of quantum models. In our case in particular, identifying inductive biases of photonic quantum circuits may help us understand where photonic QGANs may be most useful.

\section*{Acknowledgements}
We would like to thank Pierre-Emmanuel Emeriau for fruitful discussions and feedback on the manuscript. This work has been co-funded by the European Commission as part of the EIC program under the grant agreement 101130384 for the Quondensate project. We also acknowledge funding from the Plan France 2030 through the project OQuLus ANR-22-PETQ-0013.

\bibliographystyle{apsrev4-1}
\bibliography{ref}

%\clearpage
\appendix
\newpage
%\onecolumn
\onecolumngrid
\section{Algorithms}
\label{sec:algorithms}

\begin{algorithm}
    \KwData{Neural networks $G$ and $D$, data generating distribution $p_{data}(\bm{x})$}
    \KwResult{$G$ capable of sampling from $p_{data}(\bm{x})$}

    \For{number of training iterations}{
        \For{k steps}{
            Sample a batch of noise samples  $\bm{z} \sim p_z(\bm{z})$ \\
            Sample a batch of real data $\bm{x} \sim p_{data}(\bm{x})$ \\
            Generate fake data batch $G(z)$
            Get predictions for fake and real data batches: $D(G(z))$ and $D(x)$ \\
            Calculate and maximize $L_D$ according to \ref{eq:gan-final-loss}
            Update parameters of $D$
        }

        Sample a batch of noise samples  $\bm{z} \sim p_z(\bm{z})$ \\
        Get predictions for the fake data batch $D(G(z))$ \\
        Calculate and maximize $L_G$ according to \ref{eq:gan-final-loss} \\
        Update parameters of $G$
    }

    return $G$
    \caption{GAN Training}
    \label{alg:gan}
\end{algorithm}

\begin{algorithm}
    \KwData{Sub-generator count $c$, sub-generator parameters $params$ \\ \hspace*{0.95cm} Noise batch $z$, input Fock state $\ket*{\psi}$, measurement shot count $m$}
    \KwResult{A batch of fake images}

    \bigskip
    Initialize an empty batch of fake images $fakebatch$ \\
    Initialize $c$ sub-generator circuits according to $params$\\
    \ForEach{noise sample $z_i$ in $z$}{
        Initialize an empty fake image $fakeimage$ \\
        \ForEach{sub-generator $g$}{
            Encode $z_i$ into $g$ through noise reuploading phase-shifter layers \\
            Run the circuit $g$ with input $\ket*{\psi}$, performing $m$ measurement shots \\
            Build the discrete distribution of the output Fock states \\
            Map the distribution to an integer distribution. \\
            Construct the patch $g(z)$ by renormalizing the integer distribution to the interval $[0, 1]$. \\
            Add $g(z)$ to $fakeimage$
        }
        Add $fakeimage$ to $fakebatch$
    }

    return $fakebatch$
    \caption{Patch-based image generation using photonic quantum circuits}
    \label{alg:photonic-qgan}
\end{algorithm}

\section{Model details}
\label{sec:appendixmodel}

In section \ref{trainingsec} of the main text, we optimize the model in ideal conditions before proceeding with noisy simulations and the experiment. This optimization is done through hyperparameter search, by testing and choosing the models that showed the best potential and performance. Batch size is one of the fixed hyperparameters over our search. All generators are supplied $4$ noise samples and therefore produce a batch of $4$ results over which averaging is done for loss evaluation. This batch size was found to be optimal in terms of computation time and implications on the stability of the training. Another training hyperparameter which is fixed across all models is the learning rate, for both the generator and the discriminator. %The motivation behind fixing these specific parameters is to fix computational time across the models and compare their performance under the same optimization routine restrictions. This allows us to compare the models more fairly and concentrate on highlighting the most efficient quantum ansatz. 
For the discriminator, we chose a constant learning rate of $0.002$. For the generator, the adaptive learning rate of SPSA is employed, with the initial rate dependent on the initial gradient values. However, the number of optimization steps per each discriminator step is fixed to be $7$. A total of $10500$ SPSA iterations are used, with $7$ steps corresponding to a single step of gradient ascent of the discriminator. This means that the total optimization routine has $10500/7 = 1500$ steps.

Much like the baseline classical generator discussed in the main body in the assessment of the models, the discriminator is a simple fully connected network, but it has more layers and parameters. With 3 linear layers interleaved with ReLU activation functions and around 5000 parameters initialized randomly, it motivates the faster learning of the generator, by being more accurate in the classification of fake and real images. Since this is a classical model the optimisation is performed using stochastic gradient descent.

%Variable metaparameters are what makes the configurations different from each other. 
Four hyperparameters were picked to be variable: the number of sub-generators, their circuit structure as shown in Figure \ref{fig:qgan-circuits} of the main text, their input state, and the PNR capabilities of the detectors. We do a grid search on these configurations, and we present the most interesting results in Figure \ref{fig:bestconf} from the main text. The full list of the results is presented in the companion Github repository (\url{https://github.com/Quandela/photonic-qgan}) for curious readers.

From our results, we observe that deeper circuits do not necessarily mean improved results. While there is certainly an improvement when increasing the number of layers from 3 to 5, this trend does not uphold for deeper circuits. As a matter of fact, the deepest circuit from the Figure \ref{fig:bestconf-B} performs worse than the shallower analogs. This might be attributed to the difficulties associated with the training of more complex quantum circuits.

We can also see how the number of noise encoding layers affects the diversity of the generated results. Since the encoding layers are the interface for supplying the classical noise, increasing their amount means increasing the classical noise on the circuits. %This noticeably affects the diversity of the generated images, as expected. %since with more noise models produce results different from each other. 
This has an adverse effect, however, wherein the quality of the results deteriorates with the number of encoding layers. The transition between Figures \ref{fig:bestconf-C} and \ref{fig:bestconf-D} indicates a noticeable decrease in image quality with the introduction of an additional encoding layer, in otherwise identical setups. Luckily, even one encoding layer provides enough diversity in the images. The results generated by such models with one encoding layer (Figures \ref{fig:bestconf-A} and \ref{fig:bestconf-C}) may initially seem less diverse, because they preserve the same shape for all generated samples. But results are indeed diverse, as can be seen through the placements of intensity value accents in generated samples. Most importantly, simple checks show that the generated results do not replicate original training samples, but rather generate previously unseen "0"s, which means that these models generalize well.

Finally, we note that our models benefit from the availability of PNR detectors. In otherwise identical setups in Figure \ref{fig:bestconf-A}, PNR detectors make it possible to generate similar results by exploiting fewer modes and more photons, as expected. 
On the other hand, if the number of modes and photons is increased while PNR is made available, this could also potentially open up some prospects for larger-scale models and working with higher resolution images.

\newpage

\end{document}